\documentstyle[12pt,twoside,fleqn,espcrc1,psfig]{article}

\title{QCD critical point and event-by-event fluctuations
in heavy ion collisions}

\author{M.A. Stephanov\address{Institute for Theoretical Physics,
	SUNY, Stony Brook, NY 11794-3840}}

\begin{document}
\maketitle

\begin{abstract}
A summary of work done in collaboration with K. Rajagopal
and E. Shuryak. We show how heavy ion collision experiments,
in particular, event-by-event fluctuation measurements, can lead to
the discovery of the critical point on the phase diagram of QCD. 
\end{abstract}

\section{Introduction}
\begin{figure}[hbt]
\psfig{file=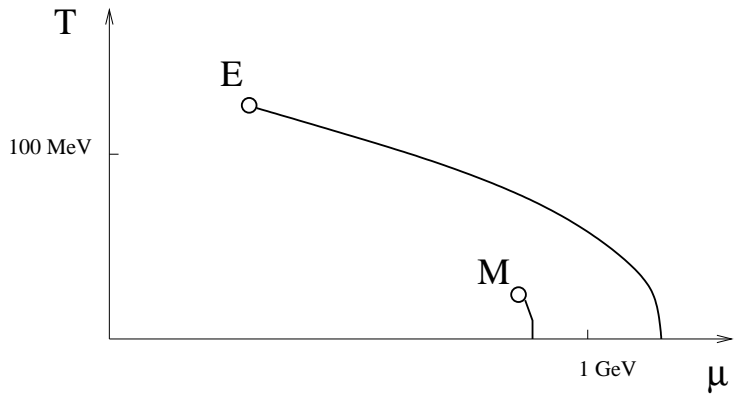,height=2in}
\caption[]{A ``minimal'' phase diagram of
QCD~\cite{SRS}. Point E is the critical end-point of the first order
phase transition separating quark-gluon and hadronic
phases.\label{fig:pd} Only the basic features relevant to this work
are indicated, a much
more complicated phase diagram is possible.}
\end{figure}

The goal of this work is to motivate a program of heavy ion collision
experiments aimed at discovering an important qualitative feature of
the QCD phase diagram --- the critical point at which a line of
first order phase transitions separating quark-gluon plasma from
hadronic matter ends~\cite{SRS} (see Fig.~\ref{fig:pd}). 
The possible existence of
such an endpoint E has recently been emphasized and its universal
critical properties have been described~\cite{BeRa97,HaJa97}. The
point E can be thought of as a descendant of a tricritical point in
the phase diagram for 2-flavor QCD with {\it massless} quarks.  As
pointed out in~\cite{SRS}, observation of the signatures of freezeout
near E would confirm that heavy ion collisions are probing above the
chiral transition region in the phase diagram.  Furthermore, we would
learn much about the qualitative landscape of the QCD phase diagram.

The basic ideas for observing the critical endpoint
proposed in~\cite{SRS} are based on the fact that such a point is a
genuine thermodynamic singularity at which 
susceptibilities diverge and the order parameter
fluctuates on long wavelengths. The resulting
signatures all share one common property: they are {\em
nonmonotonic\/} as a function of an experimentally varied parameter
such as the collision energy, centrality, rapidity or 
ion size. The goal of \cite{SRS2} is to develop a set of tools which 
will allow heavy ion collision experiments to discover the critical
endpoint through the analysis of the variation of
event-by-event fluctuations as control parameters are varied.

\section{Non-critical fluctuations and comparison with data.}

Before we can achieve our goal we must develop
sufficient understanding of {\em non-critical} event-by-event
fluctuations.
Large acceptance detectors, such as NA49 and WA98 at CERN,
have made it possible to measure important average
quantities in single heavy ion collision events, such as,
for example, the mean transverse momentum 
of the charged particles in a single event.
The most remarkable property of the data is that 
the event-by-event 
distributions of such observables are as perfect 
Gaussians as the data statistics allow.

Our first step is to analyze the NA49 data and compare it with
thermodynamic predictions for non-critical fluctuations. 
We find that the data is broadly
consistent with the hypotheses that most of the fluctuations
are thermodynamic in origin, and that PbPb collisions at 160 AGeV do not
freeze out near the critical point. This allows us to establish the
background, on top of which the effects of critical
fluctuations should be sought as the control parameters are varied.

Most of our analysis is applied to the fluctuations of the observables
characterizing the multiplicity and momenta of the charged pions in
the final state of a heavy ion collision.
We model the hadronic matter at freeze-out by a resonance gas in
thermal equilibrium. 
Our simulation 
\cite{SRS2} shows that more than half of all observed pions come from
resonance decays.  The resonances also have a dramatic effect on the
size of the multiplicity, $N$, fluctuations. We find: 
\begin{equation}\label{dN2/N} \frac{\langle (\Delta
N)^2\rangle}{\langle N\rangle} \approx 1.5 \ , 
\end{equation}
which is larger than the ideal gas value of 1.
The contribution of resonances is important to bring
this number up. The experimental value from NA49 of this ratio  is 2.0.
There is clearly room for non-thermodynamic
fluctuations, such as fluctuations of impact parameter. Their
effect can be studied and separated by varying the centrality
cut using the zero degree calorimeter.

Fluctuations of {\em intensive} observables, such as mean $p_T$ are less
sensitive to impact parameter fluctuations. However, the effects
of the flow on $p_T$ are large and complicate the analysis. 
The quantity we compare with the data is the ratio:
$v_{\rm inc} (p_T)/\langle p_T\rangle$, of the variance of the
inclusive distribution to all-event mean $p_T$. The effects
of the flow, which we do not calculate,
should largely cancel in this ratio. We find:
\begin{equation}
{v_{\rm inc} (p_T)\over\langle p_T\rangle} = 0.68.
\end{equation}
The experimental value obtained from NA49 data is 0.75.
We see that the major part of the observed fluctuation in $p_T$ 
is accounted for by the thermodynamic fluctuations. 
A large potential source of the discrepancy
is the ``blue shift'' approximation we used and could be remedied
by a better treatment of flow.

A very
important feature in the data is 
the value of the ratio 
of the scaled  event-by-event variation to the variance of the 
inclusive distribution:
\begin{equation}
F = {\langle N\rangle v^2_{\rm ebe}(p_T)\over v^2_{\rm inc}(p_T)} 
= 1.004 \pm 0.004.
\label{F}
\end{equation}
This is a remarkable fact, since the contribution of the Bose
enhancement to this ratio
is almost an order of magnitude larger than the statistical uncertainty.
Some mechanism must compensate for the Bose
enhancement. In the next section we find
a possible origin of this effect:
anti-correlations due to energy conservation and
thermal contact between the observed
pions and the rest of the system at freeze-out.

\section{Energy Conservation and Thermal Contact}
\label{contact}

We consider the effect of the energy conservation and thermal contact
between the subsystem we observe, which we call B and which consists
mainly of charged pions, and
the remaining unobserved part of the system, which we call A and 
which includes the
neutral pions, the resonances, the pions not in the experimental
acceptance and, if the freeze-out occurs near critical point, the
order parameter or sigma field. We quantify the
effect by calculating the ``master correlator'':
\begin{equation}\label{dndncorr}
\langle \Delta n_p \Delta n_k\rangle = {v}^2_p \delta_{pk}
- 
{{v}^2_p \epsilon_p {v}^2_k \epsilon_k \over 
T^2(C_A + C_B)}\ ,
\end{equation}
where $n_p$ are the pion momentum mode occupation numbers,
$v^2_p=\langle n_p\rangle(1+\langle n_p\rangle)$, and
$C_{A,B}$ are the heat capacities of the two systems A and B.

Using this expression for the correlator we can now calculate the
effect of thermal contact and energy conservation on fluctuations
of various observables, such as mean $p_T$, for example.
In particular, we find that the anti-correlation introduced by this
effect reduces the value of the ratio $F$ defined in (\ref{F})
by and amount comparable to the Bose enhancement effect, and thus
can compensate it.
This effect can be distinguished from other
effects, e.g., finite two-track resolution, also countering
the Bose enhancement, by the specific form of the
microscopic correlator (\ref{dndncorr}).  The effect of energy
conservation and thermal contact introduces an {\em off-diagonal} (in
$pk$ space, and also in isospin space)
anti-correlation. Some
amount of such anti-correlation is indeed observed in the NA49
data.
Another important point of (\ref{dndncorr}) is that as the freeze-out
approaches the critical point and $C_A$ becomes very large the
anti-correlation due to energy conservation disappears.

\section{Pions Near the Critical Point: Interaction with the Sigma Field}

\begin{figure}[t]
\centerline{\psfig{file=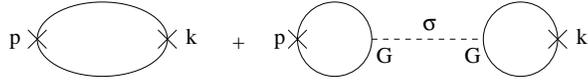,height=.4in}}
\smallskip
\caption[]{Diagrammatic representation of the right hand side
of the correlator
(\ref{dndnsigma}).}
\label{fig:diagrams}
\end{figure}

Finally, in this section, unlike the previous sections, we consider
the situation in which the freeze-out occurs very close to the
critical point. This point is characterized by large long-wavelength
fluctuations of the sigma field (chiral condensate). We must 
take into account the effect of the $G\sigma\pi\pi$ interaction
between the pions and such a fluctuating field. We do this by
calculating the contribution of this effect to the ``master
correlator''. We find:
\begin{equation}\label{dndnsigma}
\langle{\Delta n_p\Delta n_k}\rangle = {v}^2_p\delta_{pk}
+ {1\over m_\sigma^2}{G^2\over T}
{{v}^2_p{v}^2_k\over \omega_p \omega_k}.
\end{equation}
We see that the exchange of quanta of the soft sigma field (see
 Fig.~\ref{fig:diagrams}) leads to a dramatic
off-diagonal correlation, the size of which grows as we approach
the critical point and $m_\sigma$ decreases. This correlation
takes over the off-diagonal anti-correlation discussed in the
previous section.

To quantify the effect of this correlation we computed
the contribution to the ratio $F$ (\ref{F}) from (\ref{dndnsigma}).
We find:
\begin{equation}\label{FTresultmu0}
\Delta F_\sigma =  0.14 \left(G_{\rm freeze-out}\over 300\ {\rm MeV}\right)^2 
\left(\xi_{\rm freeze-out}\over 6\ {\rm fm}\right)^2 
\qquad{\rm for~} \mu_\pi = 0\ ,
\end{equation}
This effect, similarly to the Bose enhancement, is sensitive to
over-population of the pion phase space characterized by $\mu_\pi$ and
increases by a factor 2.5 for $\mu_\pi=60$ MeV.  We estimate the size
of the coupling $G$ to be around 300 MeV near point E, and the mass
$m_\sigma$, bound by finite size effects, to be less than 6 fm. The
effect (\ref{FTresultmu0}) can thus easily exceed the present statistical
uncertainty in the data (\ref{F}) by 1-2 orders of magnitude.

It is also important to note that we have calculated the effect of
critical fluctuations on $F$ because this ratio is being measured in
experiments, such as NA49. This observable is not optimized for
detection of critical fluctuations. 
Observables which are more sensitive to small $p_T$ than $F$ (e.g., ``soft
$F$''), and/or observables which are sensitive to {\em off-diagonal}
correlations in $pk$ space would show even larger effect as the
critical point is approached.


\begin{thebibliography}{9}
\bibitem{SRS} M. Stephanov, K. Rajagopal, E. Shuryak,
Phys. Rev. Lett. {\bf 81} (1998) 4816.
\bibitem{BeRa97} J. Berges and K. Rajagopal, Nucl. Phys. {\bf B538} (1999)
215.

\bibitem{HaJa97} M. A. Halasz, A. D. Jackson, R. E. Shrock, M. A. Stephanov
and J. J. M. Verbaarschot, Phys. Rev. {\bf D58} (1998) 096007.


\bibitem{SRS2} M. Stephanov, K. Rajagopal, E. Shuryak,
		hep-ph/9903292.
\end{thebibliography}
\end{document}